\documentclass[preprint,nofootinbib,aps,superscriptaddress,eqsecnum]{revtex4-1} 
 \pdfoutput=1
\usepackage[T1]{fontenc} 
\usepackage{slashed}

\def\be{\begin{equation}}
\def\ee{\end{equation}}

\def\ba{\begin{eqnarray}}
\def\ea{\end{eqnarray}}
\def\beas{\begin{eqnarray*}}
\def\eeas{\end{eqnarray*}}
\newcommand{\beq}{\begin{equation}}
\newcommand{\eeq}{\end{equation}}
\newcommand{\bea}{\begin{eqnarray}}
\newcommand{\eea}{\end{eqnarray}}
\def\nn{\nonumber}
\usepackage{graphicx}
 \usepackage{amsmath}
\usepackage{amsfonts}
\usepackage{amssymb,subfigure}
\usepackage{longtable}
\usepackage{color}
\usepackage{comment}

\begin{document}
\title{Supergravity Model of Inflation and Explaining IceCube HESE Data via PeV Dark Matter Decay}

\author{Girish Kumar Chakravarty}\email[Email: ]{girish@ctp-jamia.res.in}
\affiliation{Centre for Theoretical Physics, Jamia Millia Islamia, New Delhi, India}

\author{Najimuddin Khan}\email[Email: ]{khanphysics.123@gmail.com}
\affiliation{School of Physical Sciences, Indian Association for the Cultivation of Science, Kolkata, India}


\author{Subhendra Mohanty}\email[Email: ]{mohanty@prl.res.in}
\affiliation{Theoretical Physics Division, Physical Research Laboratory, Ahmedabad, India \vspace{1.50cm}}

\begin{abstract}
\vspace{.5cm}
We construct an unified model of inflation and PeV dark matter with an appropriate choice of no-scale K\"ahler potential, superpotential and gauge kinetic function in terms of MSSM fields and hidden sector Polonyi field. The model is consistent with the CMB observations and
can explain the PeV neutrino flux observed at IceCube HESE.
A Starobinsky like Higgs-sneutrino plateau inflation is obtained from the $D$-term SUGRA potential while $F$-term being subdominant during inflation. To get PeV dark matter, SUSY breaking at PeV scale is achieved through Polonyi field. This sets the scale for soft SUSY breaking parameters $m_0, m_{\frac{1}{2}}, A_0 $ at the GUT scale in terms of the parameters of the model.
The low energy particles spectrum is obtained by running the RGEs. We show that the $\sim$125 GeV higgs and the gauge coupling unification can be obtained in this model. The $6$ PeV bino-type dark matter is a subdominant fraction ($\sim 11\%$) of the relic density and its decay gives the PeV scale neutrino flux observed at IceCube by appropriately choosing the couplings of the $R$-parity violating operators. Also we find that there is a degeneracy in scalar field parameters $(\gamma, \beta)$ and coupling $\zeta$ value in producing the correct amplitude of CMB power spectrum. However the value of parameter $\tan(\beta)=1.8$, which is tightly fixed from the requirement of PeV scale SUSY breaking, removes the degeneracy in the values of the scalar field parameters to provide a unique solution for inflation. In this way it brings the explanation for dark matter, PeV neutrinos and inflation within the same framework.
\end{abstract}


 
\renewcommand*{\thefootnote}{\fnsymbol{footnote}}
\maketitle

 \section{Introduction}
\label{sec:dm}
The  125 GeV Higgs boson found at the Large Hadron Collider (LHC)~\cite{Aad:2012tfa,Giardino:2013bma,Chatrchyan:2012xdj}  completes the spectrum of the Standard Model but raises the question about what protects the mass of the higgs at the electroweak scale despite quantum corrections (the gauge hierarchy problem). Supersymmetry  ~\cite{Drees:2004jm,Martin:1997ns,SUSY2} has been  widely accepted as the natural symmetry argument for protecting the Higgs and other scalar masses against radiative corrections and in addition provides the unification of the gauge couplings at the GUT scale and WIMP dark matter. The idea of naturalness in SUSY~\cite{natural, Barbieri:1987fn} requires that to explain the higgs mass, loop corrections should not be too large compared to the tree level and should not rely on the cancellation of large corrections from different loop corrections. This puts upper bounds on the the masses of squarks, gluons and gauginos for natural SUSY~\cite{Barbieri:1987fn}. The searches for SUSY particles at LHC by ATLAS~\cite{ATLAS} and CMS~\cite{CMS} at $13$ TeV and with $15 ~{\rm fb^{-1}}$ of data do not find the SUSY partners and rule out the simplest models of natural SUSY ~\cite{Buckley:2016kvr,Feng:2013pwa}. The idea of supersymmetry for explaining the Higss mass without fine tuning can be abandoned while still retaining some of the positive features like coupling constant unification and WIMP dark matter in Split-SUSY~\cite{Giudice:2004tc} scenarios where squarks and gluinos are heavy evading the LHC bounds and the electro-weakinos are of the TeV scale providing the WIMP dark matter and coupling constant unification. With the tight bounds from the direct detection experiments~\cite{Aprile:2017iyp, Akerib:2016vxi} the bino-higgsino dark matter of a few $100$ GeV is getting increasingly difficult \cite{Athron:2017qdc,Abdughani:2017dqs,Zhu:2016ncq,Peiro:2016ykr,Wang:2015rli,Chakraborti:2017dpu,Abdallah:2015hza,Badziak:2017the,Cao:2016cnv,Basirnia:2016szw}. The evidence that there is need for new physics at PeV scale comes from the IceCube's High Energy Starting Events (HESE)  observations~\cite{Schneider:2019ayi, Aartsen:2013jdh,Aartsen:2013bka,Aartsen:2014gkd} of PeV energy neutrinos.
The non-observation of neutrinos with deposited energy between 0.4-1 PeV and at Glashow resonance energies has called for an extra source of PeV energy neutrinos, and a popular scenario \cite{Murase:2014tsa,Feldstein:2013kka,Sui:2018bbh,Kachelriess:2018rty,Bhattacharya:2019ucd,Pandey:2019wfk} is the decay of PeV scale dark matter with a sizable branching to neutrinos and with a lifetime of $\sim 10^{15} $ sec~\cite{Anchordoqui:2015lqa} and $\sim 10^{28} $ sec~\cite{Abbasi:2011eq,Chianese:2016smc,Roland:2015yoa,Borah:2017xgm,Aisati:2015vma,Shakya:2015tza,Dev:2016qbd,Bai:2013nga,Ahlers:2015moa,Hiroshima:2017hmy,Esmaili:2013gha,Esmaili:2015xpa,Bhattacharya:2017jaw, Boucenna:2015tra}.
Leptoquarks have been used to explain IceCube PeV events in Refs.~\cite{Anchordoqui:2006wc,Dutta:2015dka,Dey:2015eaa,Mileo:2016zeo,Barge:2016uzn,Chauhan:2017ndd}. A PeV scale supersymmetry model with gauge coupling unification, light higgs (with fine tuning) and PeV dark matter was introduced in~\cite{Wells:2004di,Ellis:2017erg}. It was found~\cite{Roland:2015yoa} that in order to obtain the large decay time of $\sim 10^{28}$ sec as required from the observed IceCube neutrino flux, dimension $6$ operators suppressed by the GUT scale had to be introduced in the superpotential. 

A different motivation for Supersymmetry is its usefulness in inflation. The low upper bound on the tensor-to-scalar ratio by Planck~\cite{Akrami:2018odb, Ade:2015xua, Ade:2015lrj} and BICEP2/Keck Array (BK15)~\cite{Ade:2018gkx, Array:2015xqh} rules out the standard particle physics models with quartic and quadratic potentials. One surviving model is the Starobinsky $R + R^2$ model  which predicts a very low tensor-to-scalar ratio of $\sim 10^{-3}$. It was shown by Ellis et al~\cite{Ellis:2013xoa} that by choosing the K\"ahler potential of the no-scale form one can achieve Starobinsky type plateau inflation in a simple Wess-Zumino model.
No-scale model of inflation from $F$-term has been constructed in SO(10)~ \cite{ Garg:2015mra, Ellis:2016ipm}, SU(5)  \cite{Ellis:2014dxa},  NMSSM \cite{Garg:2017tds} and MSSM~\cite{Chakravarty:2016fin,Chakravarty:2016avd} models.
Inflation models in Supergravity with $F$-term  scalar potential were earlier considered in~\cite{AlvarezGaume:2010rt,AlvarezGaume:2011xv,Chakravarty:2014yda,Ferrara:2014kva,Chakravarty:2015yho} and with a $D$-term scalar potential in~\citep{Ferrara:2013rsa,Farakos:2013cqa,Ferrara:2014rya,Nakayama:2016eqv} .
The Higgs-sneutrino inflation along the $D$-flat directions in MSSM has also been studied in \cite{Deen:2016zfr,Nakayama:2013nya,Aulakh:2012st,Kim:2011ay,Pallis:2011ps,Antusch:2010va,Allahverdi:2006iq}.

In this paper, we construct a $D$-term inflation model with no-scale K\"ahler potential, MSSM superpotential and an appropriate choice of gauge kinetic function in MSSM fields, which gives a Starobinsky like Higgs-sneutrino plateau inflation favored by observations. The SUSY breaking scale is a few PeV which provides a PeV scale bino as dark matter. A fraction of thermal relic density is obtained by turning on a small $R$-parity violation to give a decaying dark matter whose present density and neutrino flux at IceCube is tuned by choosing the $R$-parity violating couplings. Superysmmetry breaking is achieved by a hidden sector Polonyi field which takes a non-zero $vev$ at the end of inflation. The gravitino mass is a few PeV which sets the SUSY breaking scale. The mSUGRA model has only five free parameters including soft SUSY breaking parameters. These are the common scalar mass $m_0$, the common gaugino mass $m_{\frac{1}{2}}$, the common trilinear coupling parameter $A_0$, the ratio of Higgs field $vevs$ tan$\beta$ and the sign of mass parameter $\mu$, all are given at the gauge coupling unification scale. The spectra and the couplings of sparticles at the electroweak symmetry breaking scale are generated by renormalization group equations (RGEs) of the above soft breaking masses and the coupling parameters. The sparticle spectrum at the low energy scale was generated using the publicly available softwares {\tt FlexibleSUSY}~\cite{Athron:2014yba,Athron:2016fuq}, {\tt SARAH}~\cite{Staub:2015kfa,Staub:2017jnp} and {\tt SPheno}~\cite{Porod:2003um,Porod:2011nf} with the mSUGRA input parameters set $m_0$, $m_{\frac{1}{2}}$, $A_0$, tan$\beta$ and sign($\mu$). In our analysis, we have used {\tt SARAH} to generate model files for mSUGRA and relic density of LSP has been calculated using {\tt micrOMEGAs}~\cite{Belanger:2010gh,Belanger:2013oya}.

$R$-parity conserving SUSY models  include a stable, massive weakly interacting particle (WIPM) and the lightest supersymmetric particle (LSP), namely, neutralino which can be considered as a viable dark matter candidate. With the additional very tiny $R$-Parity violating (RPV) terms, LSP can decay to SM particles \cite{Carena:1998gd,Anchordoqui:2006wc,Dev:2016uxj,Becirevic:2018uab}, however it does not change the effective (co)annihilation cross-section appreciably in the Boltzmann equation~\cite{Ando:2015qda,Shirai:2009fq,Hamaguchi:2015wga}. 
In the context of mSUGRA model, the analysis of neutralino dark matter have been carried out in the Refs.~\cite{Baer:2008ih,Akula:2011jx,Bhattacharyya:2011se,Baer:2012uya,BhupalDev:2012ru,Kaufman:2015nda,Hu:2017hye,Baer:2003wx,Nilles:1983ge,Niessen:2008hz,Chattopadhyay:2003xi,Chan:1997bi,Baer:2015fsa}, where SUSY breaking occurs in a
hidden sector, which is communicated to the observable sector via gravitational interactions.

The paper is organized as follows. In Section~\ref{sec:model}, we display the MSSM model and the specific form of a K\"ahler potential and gauge kinetic function which gives the Starobinsky like Higgs-sneutrino plateau inflation. The $R$-parity violating terms contribute to the $F$-term potential and to the decay of the dark matter for IceCube. We fix the parameter in the gauge kinetic function and show that a $D$-term inflation model consistent with the CMB observations is obtained. In Section~\ref{sec:soft}, we describe the SUSY breaking mechanism from the Polonyi field and calculate the soft breaking parameters $m_0$, $m_{\frac{1}{2}}$, $A_0$ and gravitino mass $m_{\frac{3}{2}}$ as a function of parameters of the Polonyi potential. In Section~\ref{sec:pheno}, we give the phenomenological consequences of the PeV scale SUSY model. We show that the coupling constant unification can take place and we obtain $\sim125$ GeV Higgs mass by fine tuning. The relic density of the bino LSP is overdense after thermal decoupling but due to slow decay from the $R$-parity violation a fraction of relic density remains in the present epoch. A different $R$-parity operator is responsible for the decay of the PeV dark matter into neutrinos. We fix the parameters of the $R$-parity violating couplings to give the correct flux of the PeV scale neutrino events seen at IceCube. In Section \ref{sec:concl}, we summarize the main results and give our conclusions.

\section{$D$-term mSUGRA model of inflation}
\label{sec:model}
 
We consider the model with MSSM matter fields and $R$-parity violation and choose a K\"ahler potential $K(\phi_{i},\phi^{*}_{i},Z,Z^{*})$ and gauge kinetic fucntion $f_{ab}(\phi_{i},Z)$ with the aim of getting a plateau inflation favored by observations. 
We choose $K$ and $f_{ab}$ of the form
%
\ba
K &=&-3 \ln \Big[1 - \frac{1}{3}\Big( H^{\dagger}_{u} H_{u} + H^{\dagger}_{d} H_{d} + L^{\dagger}L + Q^{\dagger}Q  
 + \tilde e^{\ast}_R \tilde e_{R} + \tilde u^{\ast}_R \tilde u_{R} \nonumber \\&&~\hspace{3cm}+ \tilde d^{\ast}_R \tilde d_{R} 
  \Big)\Big] + Z Z^{\ast}+\frac{\alpha}{2} (Z Z^{\ast})^{2}\,, \label{KP1}  \\
f_{ab} &=& \frac{e^{-\kappa\, Z}}{1+ \zeta H_{u}\cdot H_{d}} \delta_{ab}\label{f1}  
\ea
respectively. And the MSSM superpotential $W(\phi_{i})$ 
\bea
W &=  & \mu\, H_{u} \cdot H_{d} -Y_{d}\, Q \cdot H_{d}\, \tilde d_{R} + Y_{u}\, Q \cdot H_{u}\, \tilde u_{R}  
      - Y_{e}\, L \cdot H_{d}\, \tilde e_{R} \nonumber \\&&+ \mu_{z}^{2} \, Z + \mu_{zz}\, Z^{3}
 \label{SP1}
\eea
where, 
\be
 H_{u} =\begin{pmatrix}
  \phi^{+}_{u}  \\
  \phi^{0}_{u}
 \end{pmatrix}\,,~~~
 H_{d} =\begin{pmatrix}
  \phi^{0}_{d}  \\
  \phi^{-}_{d}
 \end{pmatrix}\,,~~~
L =\begin{pmatrix}
  \phi_{\nu}\\
  \phi_{e}
 \end{pmatrix}\,,~~~
Q =\begin{pmatrix}
  u_{L}  \\
  d_{L}
 \end{pmatrix}\,,\label{LHu}
\ee 
and field $Z$ is the hidden sector Polonyi which is introduced to break supersymmetry. The Polonyi field $Z$ is associated with the fluctuations in the overall size of the compactified dimensions which has to be strongly stabilized at SUSY-breaking for the successful implementation of inflation in supergravity. The parameters $\alpha$, $\kappa$, $\mu_{z}$, $\mu_{zz}$ and $\zeta$ are coupling constants of the model to be fixed from SUSY breaking and inflation. The other fields bear their standard meanings. Also the above potentials are in the $M_{p}=(8\pi G)^{-1}=1$ unit and shall use the same convention through out the analysis of this model.

In addition to the superpotential given in Eq.~(\ref{SP1}), we also consider the $R$-parity violating interaction terms
\be
W_{int}=\lambda_{ijk}L_{i}L_{j}e_{Rk}^{c} + \lambda^{\prime}_{ijk}L_{i}Q_{j}d_{Rk}^{c}
+\frac{1}{2}\lambda^{\prime\prime}_{ijk}u_{Ri}^{c} d_{Rj}^{c} d_{Rk}^{c}
\label{SP2}
\ee
which will play role in explaining observed DM relic density and PeV neutrino flux at IceCube.

In supergravity, the scalar potential depends upon the K\"ahler function $G(\phi_{i},\phi^{*}_{i})$ given in terms of superpotential $W(\phi_{i})$ and K\"ahler potential $K(\phi_{i},\phi^{*}_{i})$ as $ G(\phi_{i},\phi^{*}_{i}) \equiv K(\phi_{i},\phi^{*}_{i}) + \ln W(\phi_{i}) +\ln W^{\ast}(\phi^{*}_{i})$, where $\phi_{i}$ are the chiral scalar superfields. 
In $\mathcal{D}=4$, $\mathcal{N}=1$ supergravity, the total tree-level supergravity scalar potential is given as the sum of $F$-term and $D$-term potentials which are given by
\be
V_{F}=e^{G}\left[\frac{\partial G}{\partial \phi^{i}} K^{i}_{j*} \frac{\partial G}{\partial \phi^{*}_{j}} - 3 \right] \label{LV}
\ee
and
\be
V_D= \frac{1}{2}\left[\text{Re}\, f_{ab}\right]^{-1} D^{a}D^{b},\label{VD1}
\ee
respectively, where $D^{a}=-g \frac{\partial G}{\partial \phi_{k}}(\tau^{a})_{k}^{l}\phi_{l}$ and $g$ is the gauge coupling constant corresponding to each gauge group and $\tau^{a}$ are corresponding generators. For $SU(2)_L$ symmetry $\tau^{a}=\sigma^{a}/2$, where $\sigma^{a}$ are Pauli matrices. And the $U(1)_{Y}$ hypercharges of the fields $H_u$, $H_d$, $L$, $Q$, $\tilde e_{R}$, $\tilde d_{R}$, $\tilde u_{R}$ given in (\ref{SP1}) are $Y = (\frac{1}{2}, -\frac{1}{2}, -\frac{1}{2},\frac{1}{6}, 1, \frac{1}{3}, -\frac{2}{3} )$ respectively. The quantity $f_{ab}$ is related to the kinetic energy of the gauge fields and is a holomorphic function of superfields $\phi_i$. The kinetic term of the scalar fields is given by
\be
\mathcal{L}_{KE}=K_{i}^{j*} \partial_{\mu}\phi^{i} \partial^{\mu}\phi^{*}_{j}~, 
\label{LK}
\ee
where $K^{i}_{j*}$ is the inverse of the K\"ahler metric $K_{i}^{j*} \equiv \partial^{2}K / \partial\phi^{i}\partial\phi^{*}_{j}$. 
We assume that during inflation, SUSY is unbroken and the hidden sector field is subdominant compared to inflaton field so that we can safely assume $Z=0$.
Also, for charge conservation, we assume that during inflation the charged fields take zero vev. The $D$-term and $F$-term potentials are obtained as
\ba
V_{D} &=& \frac{9}{8}(g_{1}^{2} + g_{2}^{2}) \frac{\left(|\phi_{d}|^2-|\phi_{u}|^2+|\phi_{\nu}|^2\right)^{2} \left(-1+\zeta |\phi_{d}|^2\right)^{2}}{\left(-3+|\phi_{d}|^2+|\phi_{u}|^2+|\phi_{\nu}|^2\right)^{2}}\,,\label{VD}\\ 
V_{F} &=& \frac{3\mu^{2} \left(3|\phi_{u}|^2+|\phi_{d}|^2(3-|\phi_{u}|^2)\right)}{\left(-3+|\phi_{d}|^2
+|\phi_{u}|^2+|\phi_{\nu}|^2\right)^{2}}\,.\label{VF}
\ea
It can be seen that the above expressions reduces to MSSM $D$-term and $F$-term potentials in the small field limit after the end of inflation as the terms coming from K\"ahler potential and gauge kinetic function are Planck suppressed. Now, for simplification, we parametrize the neutral component fields as 
\ba
&&\phi^{0}_{u} = \phi\, \sin[\beta]\,, ~~~~~~~\phi^{0}_{d}= \phi\, \cos[\beta]\,, ~~~~~~~~\phi_{\nu}=\gamma \phi \,.\label{Dflatfields}
\ea
For the above parametrization, the kinetic term turns out to be
\be
\mathcal{L}_{KE} = \frac{9(1+\gamma^{2})}{\left(-3+(1+\gamma^{2})|\phi|^2\right)^2} |\partial_{\mu}\phi|^2\,.
\ee
To obtain the canonical kinetic term for the inflaton field and to better understand the inflation potential, 
we redefine the field $\phi$ to $\chi_{c}$ via
\be
\phi = \frac{\sqrt{3}}{\sqrt{1+\gamma^2}} \tanh\left(\frac{\chi_{c}}{\sqrt{3}}\right)\,.\label{PhiToChi}
\ee
For the above field redefinition, the kinetic term becomes
\be
\mathcal{L}_{KE} = {\rm sech}^2\left(\frac{2\, {\rm Im}[\chi_c]}{\sqrt{3}}\right) |\partial_{\mu}\chi_{c}|^2\,,
\ee
therefore, if the imaginary part of the field $\chi_c$ is zero, we obtain the canonical kinetic term in real part of field $\chi_{c} =\chi$(say).
As the field $\phi$ is a linear combination of the Higgs and sneutrino field, so is the inflaton field $\chi$, we can call this model a Higgs-sneutrino inflation model. The $D$-term potential (\ref{VD}) in the canonical inflaton field $\chi$ becomes
\ba
V_{D} = \frac{1}{2}\lambda^{2}\,(g_{1}^{2} + g_{2}^{2})  \tanh^{4}\left(\frac{\chi}{\sqrt{6}}\right)\,,\label{VDpot}
\ea
 where $\lambda = \frac{3}{2} \,\frac{\gamma^{2}+ \cos(2\beta)}{1+ \gamma^{2}}$
and we have made a specific choice $\zeta =\frac{\gamma^{2}+1}{3 \sin(\beta) \cos(\beta)}$ which is critical to obtain a plateau behavior potential at large field values which can fix inflationary observables. Since at the GUT scale, the mass parameter $\mu \sim 0$ and SU(2) gauge couplings are $g_{1}=g_{2}=0.62$, the $D$-term potential dominates over $F$-term potential during inflation when the field values are at the Planck scale. With the canonical kinetic term and scalar potential obtained in canonical inflaton field $\chi$,
the theory is now in the Einstein frame. Therefore, we can use the standard Einstein frame relations to estimate the 
inflationary observables, namely, amplitude of the curvature perturbation $\Delta_{\mathcal R}^{2}$, 
scalar spectral index $n_{s}$ and its running $\alpha_s$, and tensor-to-scalar ratio $r$, given by
\ba
\Delta_{\mathcal R}^{2} &=& \frac{1}{24 \pi^{2}} \frac{V_D}{\epsilon}\,,\label{amplitude}\\
n_{s} &=& 1-6\epsilon+2\eta\,,\\
\alpha_{s} &\equiv& \frac{dn_{s}}{d\ln k} = 16\epsilon\eta -24\epsilon^{2} - 2\xi\,,\\
r &=& 16\epsilon\,,
\ea
respectively. Here $\epsilon$, $\eta$ and $\xi$ are the 
slow-roll parameters, given by
\be
\epsilon = \frac{1}{2}\left(\frac{V_{D}'}{V_{D}}\right)^2, 
~~~~~~~\eta = \frac{V_{D}''}{V_{D}}\,,
~~~~~~~\xi = \frac{V_{D}'V_{D}'''}{V_{D}^{2}}.
\ee

In order to have flat Universe as observed, the universe must expand at least by more than $60$ $e$-folds
during inflation. The displacement in the inflaton field during inflation is $\Delta \chi \equiv \chi_i - \chi_f$. 

\begin{figure}
    \centering
    \begin{minipage}{0.46\textwidth}
        \centering
        \includegraphics[width=0.9\textwidth]{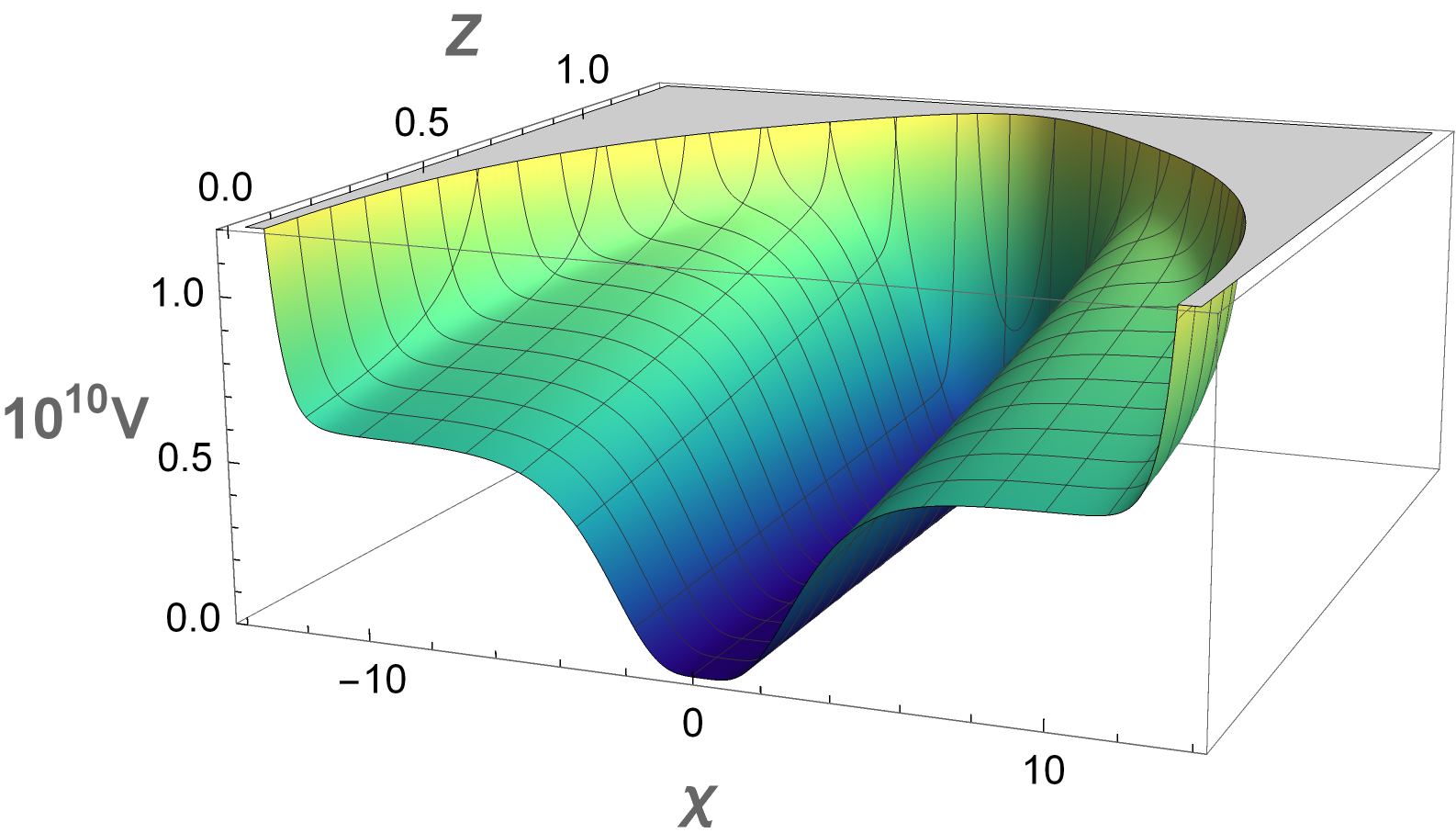} 
        \caption{\footnotesize \textit{The $D$-term inflation potential for $\tan(\beta)=1.8$, $g_{1}=g_{2}=0.62$ is shown. During inflation as the inflaton rolls down from $\chi_{i}\simeq7$ to $\chi_e\simeq1.9$, the potential along the $Z$ direction stays nearly flat implying inflaton $\chi$ being much heavier than the Polonyi field $Z$. For $Z \gtrsim 1$, the potential becomes very steep where slow-roll inflation cannot be achieved.}}
        \label{fig1}
    \end{minipage}\hfill
    \vspace{.4cm}
    \begin{minipage}{0.46\textwidth}
        \centering
        \includegraphics[width=0.9\textwidth]{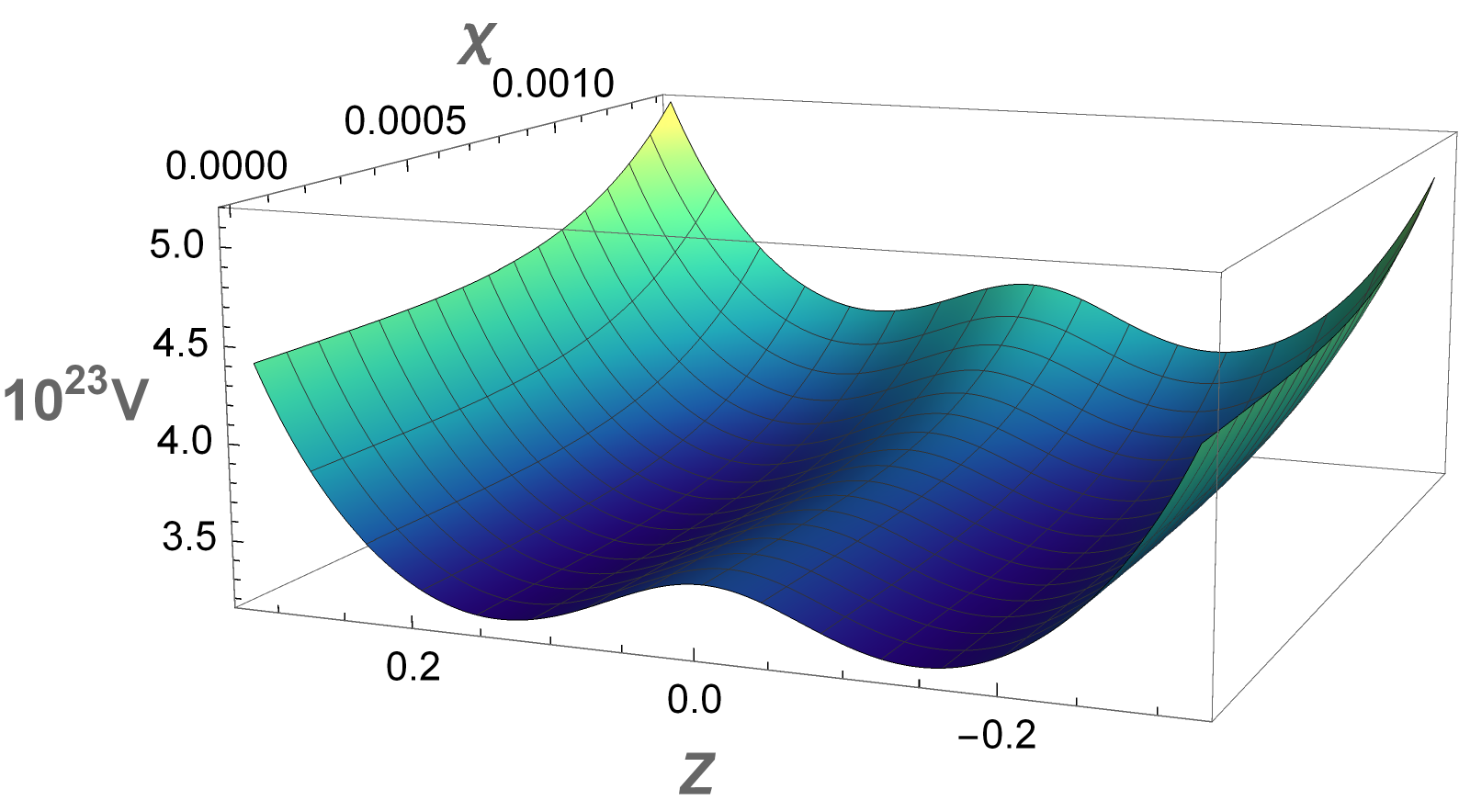} 
        \caption{\footnotesize \textit{After the end of inflation as $\chi$ approaches minimum near zero, Polonyi field $Z$ settles down to minimum at $Z_{min}\simeq \pm 0.144$. The positive Polonyi potential is obtained for the parameter vlaues $\alpha= 20$, $\mu_{z}\simeq 2.43 \times 10^{-6} $, $\mu_{zz}\simeq 2.858\times 10^{-11}$, $\kappa=0.56$, which are fixed from the requirement of getting PeV scale soft SYSY breaking parameters.}}
        \label{fig2}
    \end{minipage}
    \vspace{.4cm}
\end{figure}

The field value $\chi_i$ is at the onset of inflation, when observable CMB modes starts leaving the horizon, can be
determined using the following relation
\be
N=\int_{\chi_f}^{\chi_i} \frac{V_D}{V_{D}'} d\chi\,,\label{efolds}
\ee 
once we fix the field value $\chi_f$ using the condition $\epsilon(\chi_f)=1$ which corresponds to end of inflation. 



From the Planck-2018 CMB temperature anisotropy data in combination with the EE measurement
at low multipoles, we have the scalar amplitude, the spectral index and
its running as $\ln (10^{10} \Delta_{\mathcal R}^{2}) = 3.044\pm 0.014$, $n_{s}= 0.9649 \pm 0.0042$,
$\alpha_s = - 0.0045 \pm 0.0067$, respectively, at ($68 \%$ CL, PlanckTT,TE,EE+lowE+lensing)~ \cite{Akrami:2018odb, Ade:2015xua, Ade:2015lrj}.
Also the Planck-2018 data combined with BK15 CMB polarization data
put an upper bound on tensor-to-scalar ratio $r_{0.002} < 0.056 \,(95\%$ CL, PlanckTT,TE,EE+lowE+lensing+BK15)~\cite{Ade:2018gkx, Array:2015xqh}. 
Armed with the theoretical and observational results, we perform the numerical analysis of the models.
From the condition of end of infaltion $\epsilon=1$, we find $\chi_e\simeq1.927$. Therefore requiring $N\simeq60$ $e$-folds expansion during inflation,
we obtain $\chi_{i}\simeq7.03$.
At $\chi_{i}\simeq7.03$, CMB observables are estimated to be $r \simeq 0.0035$, 
$n_{s} \simeq 0.965$ and $\alpha_{s} \simeq -6\times10^{-4}$, consistent with the observations. The $(n_{s}, r)$ predictions are similar to  Starobinsky inflation.
For $\tan(\beta)=1.8$ and $g_{1}=g_{2}=0.62$, the observed CMB amplitude can be obtained for $\lambda\simeq 1.703\times 10^{-5}$. From the specific case considered in (\ref{VDpot}) for successful inflation, we have sneutrino field parameter $\gamma\simeq 0.727$, and coupling $\zeta$ of the Planck suppressed operator in the guage kinetic function (\ref{f1}) $\zeta\simeq 1.2$. The evolution of the $D$-term inflaton potential (\ref{VDpot}) is shown in Fig.~\ref{fig1}. During inflation as the inflaton rolls down from $\chi_{i}\simeq7$ to $\chi_e\simeq1.9$, the potential along the $Z$ direction stays nearly flat implying $\chi$ being much heavier than $Z$. During slow-roll phase, the mass of inflaton varies from $m_{\chi_{i}}\simeq 2.1\times 10^{13}$ GeV to $m_{\chi_{e}}\simeq 9.3\times 10^{12}$ GeV, whereas the mass parameter associated with the Polonyi field $Z$ varies from $m_{Z_{i}}\simeq 6.1\times 10^{10}$ GeV to $m_{Z_{e}}\simeq 2.1\times 10^{8}$ GeV. Hence, $m^{2}_{Z}\ll m^{2}_{\chi}$, our assumption that $Z$ is subdominant during inflation and contributes insignificantly to the supergravity inflation potential.

In the next Section we will see that the value of $\tan(\beta)=1.8$ used to fix CMB amplitude is absolutely critical in order to get PeV dark matter whose decay explain PeV neutrino events observed at IceCube, and reproducing low energy particles mass spectrum as shown in Table~\ref{tab2}. It is important to mention here that the scalar field parameter values $\tan(\beta)=1.8$ and $\gamma\simeq 0.727$ is not a unique set of value to obtain the observed CMB amplitude. Instead it can be obtained for all the pair of values of $(\gamma, \beta)$ which satisfy the relation $ \frac{3}{2} \,\frac{\gamma^{2}+ \cos(2\beta)}{1+ \gamma^{2}}=\lambda\simeq1.703\times 10^{-5}$ defined in (\ref{VDpot}). However, $\tan(\beta)=1.8$ fixed from the requirement of PeV scale SUSY breaking removes this degeneracy in $(\gamma, \beta)$ in obtaining the correct CMB amplitude and, at the same time, it brings the successful explanation for inflation, dark matter relic density and PeV neutrino events at IceCube within the same framework. 


\section{SUSY breaking}
\label{sec:soft}
We assumed that during inflation all the fields including the hidden sector field $Z$ are subdominant compared to inflaton field and supersymmetry is unbroken at the time of inflation. Once the inflation ends, the scalar fields 
effectively become vanishing and the soft mass terms are generated via SUSY breaking as the field $Z$ 
settles down to a finite minimum of the Polonyi potential. The Polonyi field $Z$ is associated with the fluctuations in the overall size of the compactified dimensions. 
 For the successful implementation of inflation in supergravity the strong stabilization of the Polonyi field is required which can be achieved via appropriate choice of K\"ahler potential, guage kinetic function and superpotential in $Z$. This also allows for a solution of the cosmological Polonyi problem \cite{Coughlan:1983ci, Ellis:1986zt} (which is a special case of cosmological muduli problem \cite{Banks:1993en,deCarlos:1993wie}) associated with the problem of Cosmological Nucleosynthesis. Technically, cosmological Polonyi problem is evaded if $Z$ is heavy $m_{Z}\sim \mathcal{O} (100-1000)$ TeV and the Polonyi mass is much larger than the gravitino mass, $i.e$ $m^{2}_{Z}\gg m^{2}_{3/2}$\, \cite{Linde:2011ja, Dudas:2012wi,Chakravarty:2016avd}. We will see that this problem does not occure in this model. 
 
 The appropriatly chosen potentials $K(Z,Z^{\ast})$ and $W(Z)$ are shown in eq.(\ref{KP1}) and eq.(\ref{SP1}), and the gauge kinetic function in eq.(\ref{f1}).
The kinetic term for Polonyi field $Z$ is obtained as $\left(1 + 2\alpha|Z|^{2}\right)|\partial_{\mu}Z|^{2}$ and the Polonyi potential is obtained as
\bea
V(Z, Z^{\ast}) &=& e^{|Z|^{2} + \frac{\alpha}{2} |Z|^{4}}  |Z|^{2} |\mu_{zz} Z^{2} + \mu_{z}^{2}|^{2} 
\Big[ -3 + \frac{1}{1+2 \alpha  |Z|^{2}} \Big|\frac{1}{Z} \nonumber \\&&~~~~~~+ Z^{\ast} +\alpha Z Z^{\ast 2} + \frac{2\mu_{zz} Z}{\mu_{zz} Z^{2} + \mu_{z}^{2}} \Big|^2 \Big]\,.\label{Vz}
\eea
We take the hidden sector Polonyi field $Z$ to be real. The parameters $\alpha$, $\mu_{z}$ and $\mu_{zz}$ are fixed from the requirement that $Z$ which breaks supersymmetry acquire a minima where the gravitino mass is $\sim$ $\mathcal{O}({\rm PeV})$ scale and the Polony potential is positive $V_{F}(Z) \gtrsim 0$ with a field minimum at $Z=Z_{min}$. We find gravitino mass to be
\be
m_{\frac{3}{2}}^{2}=e^{G}= e^{Z^2 + \frac{\alpha}{2}Z^4} (\mu_{zz} Z^3 +\mu_{z}^{2} Z)^{2},\label{m3/2}
\ee
and the scalar masses, using $m^{2}_{\phi} = \frac{\partial_{\phi}\partial_{\phi^{\ast}}(V_{F}(\phi,\phi^{\ast},Z))}{\partial_{\phi}\partial_{\phi^{\ast}}K}$ evaluated at $\phi=0$, $Z=Z_{min}$, are obtained as
\begin{align}
\label{mphii}
m_{\phi_{i}}^{2} &= m_{\frac{3}{2}}^{2} \left[-2 + \frac{1}{1+2\alpha Z^{2}} \left( \frac{1}{Z} + Z +\alpha Z^{3} + \frac{2\mu_{zz} Z}{\mu_{zz} Z^{2} + \mu_{z}^{2}} \right)^{2}  \right]\,,\\
\label{mphiud}
m_{\phi_{(u,d)}}^{2} &= m_{\phi_{i}}^{2} + \frac{\mu^{2}}{(\mu_{zz} Z^{3}+ \mu_{Z}^{2} Z)^{2}} m_{\frac{3}{2}}^{2} \,
\end{align}
where $\phi_{i}=\phi_{\nu}, \phi_{e}, \phi^{+}_{u}, \phi^{-}_{d}, \tilde u_{L}, \tilde d_{L}, \tilde u_{R}, \tilde d_{R}, \tilde e_{R}$.
At the GUT energy scale, $\mu \sim 0$, all the scalar masses are equal $m_{\phi_{(u,d)}}^{2} = m_{\phi_{i}}^{2}$.

Now we calculate the coefficients of the 
soft SUSY breaking terms which arise from the K\"ahler potential (\ref{KP1}) and superpotential
(\ref{SP1}). The effective potential of the observable 
scalar sector consists of soft mass terms which give scalar masses as given by Eq.s\,
(\ref{mphii})-(\ref{mphiud}), and trilinear and bilinear soft SUSY breaking terms given by\,\cite{Soni:1983rm,Kaplunovsky:1993rd,Dudas:2005vv}
\be
\frac{1}{3}A_{ijk}\phi^{i}\phi^{j}\phi^{k}
+\frac{1}{2}B_{ij}\phi^{i}\phi^{j}+ \rm{h.c.}\,,\nonumber
\ee
here the coefficients $A_{ijk}$ and $B_{ij}$ given by
\begin{subequations}
\begin{align}
\label{tri}
A_{ijk} &= \left[\frac{\partial_{Z^{\ast}}\hat{W}^{\ast} +\hat{W}^{\ast}\partial_{Z^{\ast}}\hat{K} }{\partial_{Z^{\ast}} \partial_{Z} \hat K}\,\partial_{Z}\, e^{\hat{K}}\right]\, \tilde Y_{ijk}  \,,\\
\label{bi} 
B_{ij} &= \left[\frac{\partial_{Z^{\ast}}\hat{W}^{\ast} +\hat{W}^{\ast}\partial_{Z^{\ast}}\hat{K}}{\partial_{Z^{\ast}} \partial_{Z} \hat K} \,\partial_{Z}\, e^{\hat{K}}- m_{\frac{3}{2}} e^{\hat K/2}\right] \,\tilde \mu_{ij} \,,
\end{align}
\end{subequations}
From Eqs.~(\ref{KP1}) and (\ref{SP1}), we have $\hat{K}(Z,Z^{\ast})=Z Z^{\ast}+
\frac{\alpha}{2} (Z Z^{\ast})^{2}$ and $\hat{W}(Z)= \mu_{z}^{2} Z + \mu_{zz}Z^{3}$.
The coefficients of normalised masses $\tilde \mu_{ij}$ and Yukawa couplings $\tilde Y_{ijk}$ in the trilinear (\ref{tri}) and bilinear (\ref{bi}) terms, respectively, are obtained as

\begin{align}
\label{eq:Aijk}
A_{0} &= e^{Z^2 + \frac{\alpha}{2}Z^4} \frac{Z (1+Z^{2}\alpha)\left[\mu_{zz} Z^{2} (3 + Z^{2} 
+ Z^{4}\alpha) + (1 + Z^{2} + Z^{4}\alpha)\mu_{z}^{2} \right]}{1+ 2 \alpha Z^{2}}\,,\\
\label{eq:Bij}
B_{0} &= A_{0} -e^{Z^2 + \frac{\alpha}{2}Z^4} \, (\mu_{zz} Z^{3} + \mu_{z}^{2}Z)\,.
\end{align}
We drop the $B_{ij}$ contribution to the scalar masses by taking corresponding $\tilde\mu_{ij}$ small. 
Also from the fermionic part of the SUGRA Lagrangian, the soft gaugino masses can be obtained as~\cite{Brignole:1993dj,Brignole:1997dp,Dudas:2005vv}
\be
m_{\frac{1}{2}} = \frac{1}{2} \left[\text{Re} \, f_{ab}\right]^{-1}\, e^{G/2} \,\partial_{I}f_{ab}\,\hat K^{I J^{\ast}} G_{J^{\ast}}\,.\label{mhalf}
\ee
For the choice of gauge kinetic function $f_{ab} = e^{-\kappa\, Z}\, \delta_{ab}$ when $\phi\approx 0$ after inflation, we obtain the gaugino mass
\be
m_{\frac{1}{2}} = \frac{\kappa}{2}\, m_{\frac{3}{2}} \,  (1+2\alpha Z^{2}) \left( Z+\alpha Z^{3}+ \frac{\mu_{z}^{2}+3\mu_{zz}Z^{2}}{\mu_{z}^{2} Z + \mu_{zz} Z^{3}} \right)\,.
\ee
We show the Polonyi potential (\ref{Vz}) in Fig.\ref{fig2} for the parameters $\alpha= 20$, $\mu_{z}\simeq 2.43 \times 10^{-6} $, $\mu_{zz}\simeq 2.858\times 10^{-11}$, $\kappa=0.56$ has a minimum at $Z_{min}\simeq \pm 0.144$. These parameter values are
fixed at the GUT scale from the requirement to achieve the soft SUSY breaking parameters
\be
A_0=-2.2\, {\rm PeV}\,,~~~~~ m_{0}=m_{\phi}=14\, {\rm PeV}\,,~~~~~m_{\frac{1}{2}}\simeq 10\,{\rm PeV}\,.\nn
\ee
with the specific choice of $\text{tan}\beta=1.8$ and sign of $\mu>0$ which gives the Higgs mass $\sim 125$ GeV at electroweak scale satisfing all experimetal (LHC etc) and theoretical constraits (stability, unitarity etc). The gravitino mass and the mass of the Polonyi field comes out to be $m_{\frac{3}{2}}\simeq 2.32\,{\rm PeV}$ and $m_{z}\simeq31.8\, {\rm PeV}\,$, respectively, which implies $m^{2}_{Z}\gg m^{2}_{3/2}$ and therefore the $\mathcal{O}({\rm PeV})$ scale oscillations of the Polonyi field near its minimum decay much before the Big Bang Nucleosynthesis. This leads to strong stabilization of $Z$ and  therefore cosmological Polonyi problem does not occur in this model.


Knowing the soft breaking parameters at a high energy scale does not tell us anything about the phenomenology we could observe in the experiments such at LHC, direct detection of DM, IceCube, etc. We need to find these parameters at the low energy. In general all parameters appearing in the supersymmetric Lagrangian evolve with RGEs. These RGEs are the inter-mediator between the unified theory at GUT scale and the low-energy masses and couplings, which strictly depend on the boundary conditions.
We apply RGEs to calculate low-energy masses and different branching ratios for
the above mentioned set of mSUGRA parameters. As the RGEs are coupled differential equations, which cannot be solved analytically. Also the low-energy phenomenology is complicated due to mixing angles and dependence of couplings on the high-scale parameters, therefore one has to rely on numerical techniques to solve RGEs. We calculate all the variables as allowed by the present experimental data. There are various programs publicly available such as {\tt SARAH, SPheno, Suspect} which can generate two-loop RGEs and calculate the mass spectrum and the couplings at low-energy. For this work, we use {\tt SARAH-4.11.0}~\cite{Staub:2015kfa,Staub:2017jnp} to generate the RGEs and other input files for {\tt SPheno-4.0.2}~\cite{Porod:2003um,Porod:2011nf} which generate the mass spectrum, couplings, branching ratios and decay widths of supersymmetric particles. To study the neutralino as a dark matter candidate we link the output files of {\tt SPheno} and model files from {\tt SARAH} to {\tt micrOMEGAs-3.6.8} \cite{Belanger:2010gh,Belanger:2013oya} to calculate the number density for a PeV neutralino dark matter. In the next section we discuss the consequences of a PeV dark matter.

\section{Bino dominated DM in mSUGRA model}
\label{sec:pheno}
\begin{figure}[h!]
 \begin{center}
 \includegraphics[width=3.8in,height=2.8in, angle=0]{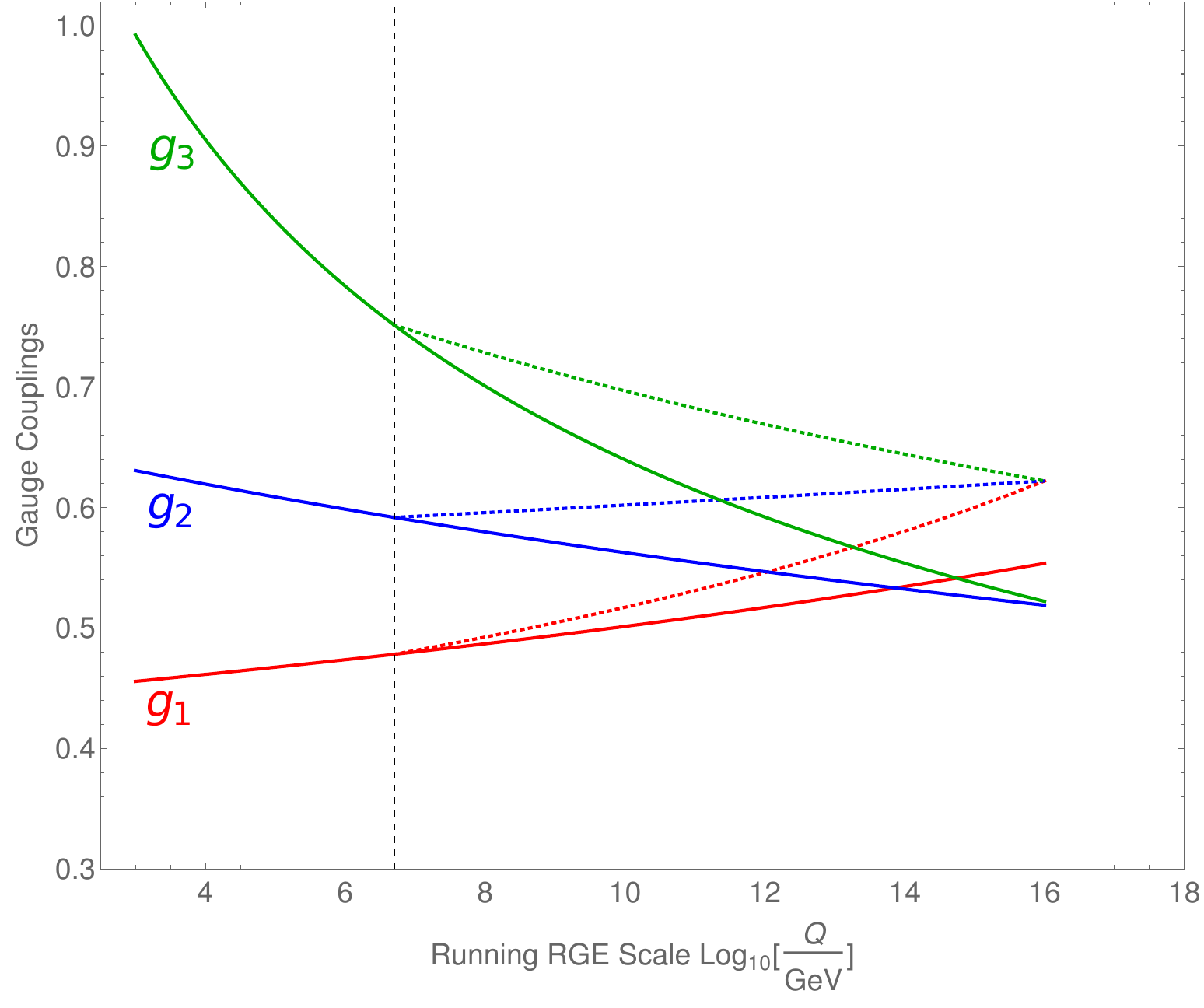}
 \caption{\label{fig:gaugeUni}\textit{Solid lines represent the gauge coupling evolution for the standard model.
Dotted lines represent the gauge coupling unification for our benchmark points (Table~\ref{tab1}).} }
 \end{center}
 \end{figure}

The neutralinos $\chi_i$ (i=1,2,3,4) are the physical superpositions of two gauginos namely bino $\tilde{B}$ and wino $\tilde{W}_3$, and two Higgsinos $\tilde{H}_{u0}$ and $\tilde{H}_{d0}$. The neutralino mass matrix is given by,
\begin{equation}
M_N =\begin{pmatrix}
 M_1  &  0 & -M_Z c_b s_\theta & M_Z s_b s_\theta \\
 0    & M_2 & M_Z c_b c_\theta & -M_Z s_b c_\theta \\
 -M_Z c_b s_\theta & M_Z c_b c_\theta&0&-\mu\\
 M_Z s_b s_\theta & -M_Z s_b c_\theta&-\mu&0\\
\end{pmatrix} \nonumber
\end{equation}
where $c_b\equiv \text{cos}\beta$, $c_\theta\equiv \text{cos}\theta_W$ and $\theta_W$ is the Weinberg mixing angle. $M_1$ and $M_2$ are the bino and wino mass parameter at the EWSB scale. The lightest eigenvalue of the above matrix and the corresponding mass eigenstate has good chance of being the LSP. The Higgsinos dominantly contribute in LSP for $|\mu|<M_{1,2}$, whereas for $|\mu|>M_{1,2}$, the LSP can be determined by bino and wino. The lightest neutralino becomes bino-like when $M_2 > M_1$. In mSUGRA model, $M_1$ and $M_2$ are equal due to the universality of the gaugino masses at the GUT scale where the value of the gauge coupling constants $g_{1}$, $g_{2}$ and $g_{3}$ become equal, see Fig.~\ref{fig:gaugeUni}. However at low energy $M_2 \approx \frac{5 g^2_2 (M_Z)}{3 g^2_1 (M_Z)} M_1$~\cite{Niessen:2008hz}. This implies that the LSP is mostly bino-like. 

\begin{table}[h]
\begin{center}
\begin{tabular}{|c|c|c|c|c|c|c|}
\hline
\hline
 \multicolumn{4}{|c|}{SUGRA parameters}&\multicolumn{1}{c|}{ RPV}& DM mass & {Decay time }\\
 \cline{1-4}
 \multicolumn{3}{|c|}{Masses in PeV}&&couplings & in PeV & {in (secs), }\\
 \multicolumn{3}{|c|}{}&&&  & { Explains }\\
\cline{1-3} \cline{5-5}   
  & &&& & &  IceCube-\\
 $m_0$ &$m_{\frac{1}{2}}$ &$A_0$ & $\tan\beta$ & $\lambda_{121}$ && HESE \\
&&&&  &&  events  \\
\hline
\hline
&&&&&&\\
 14&10&-2.2& 1.8& $2.82\times 10^{-28}$ & 6.3 &$8.90\times 10^{28}$\\
 &&&&&&\\
\hline
\hline
\end{tabular}
\end{center}
\caption{\textit{A list of the Benchmark points (BP) as used in our analysis. Using these BPs, we have reproduced the Higgs mass at $\sim$125 GeV and explain the relic density and the PeV neutrino events seen at the IceCube experiment.}}
\label{tab1}
\end{table}
\begin{table}[h]
\begin{center}
\resizebox{12cm}{!}{
\begin{tabular}{cccc}
\hline \hline 
SUSY Fields &  & ~Masses in PeV& \\ 
\hline \hline 
\(\tilde{d}_{iL,R}\)~($i=d,s,b$) & &$M_{\tilde{d}_{i,L,R}}\approx$ $(15.75,~16.78)_d$~$(16.78,~16.78)_s$~$(17.38,~17.38)_b$ \\ 
  \hline 
 \(\tilde{u}_{iL,R}\)~($i=u,c,t$)  & &$M_{\tilde{u}_{iL,R}}\approx$ $(13.30,~15.75)_u$~$(16.88,~16.88)_c$~$(17.38,~17.38)_t$\\ 
  \hline 
 \(\tilde{e}_{iL,R}\)~($i=e,\mu,\tau$)  & &$M_{\tilde{e}_{iL,R}}\approx$ $(14.36,~14.36)_e$~$(14.36,~14.86)_\mu$~$(14.86,~14.86)_\tau$\\ 
   \hline 
 \(\tilde{\nu}_i\)~~~($i=e,\mu,\tau$) & & $M_{\tilde{\nu}_i}\approx$ $(14.86)_e$~~~~$(14.86)_\mu$~~~~$(14.86)_\tau$\\ 
\hline
 \(h\)~~~~~~~~~~~~~~~ & & $M_h\approx126$ [GeV]\\ 
  \hline
 \(H\)~~~~~~~~~~~~~~~ & & $M_H\approx 19.48$\\ 
  \hline 
 \(A\)~~~~~~~~~~~~~~~ & & $M_{A}\approx 19.48$\\ 
  \hline 
 \(H^-\)~~~~~~~~~~~~~~ & & $M_{H^-}\approx 19.47$\\ 
 \hline 
\(\tilde{g}_i\) ~~~($i=1...8$)& & $M_{\tilde{g}_i}\approx 14.28$\\ 
 \hline  
 \(\tilde{\chi}^0\) ~~~($i=1...4$) & & $M_{\tilde{\chi_1}} (DM)\approx 6.3$~~~$M_{\tilde{\chi_2}}\approx7.22$~~~~$M_{\tilde{\chi_3}}\approx7.22$~~~~$M_{\tilde{\chi_4}}\approx9.22$\\ 
  \hline 
 \(\tilde{\chi}^-\)~~~~~~~~~~~~~~~ & & $M_{\tilde{\chi}_1^-}\approx7.22$~~~~$M_{\tilde{\chi}_2^-}\approx9.21$\\ 
 \hline
\end{tabular}}
\end{center}
\caption{\textit{Particle mass spectrum after EWSB for the benchmark point given in Table~\ref{tab1}.}}
\label{tab2}
\end{table}
\subsection{Relic density and PeV excess at IceCube}
In this analysis we present our mSUGRA model with the aim to
realize a PeV scale dark matter candidate, which can also explain the IceCube HESE events.
IceCube~\cite{Schneider:2019ayi, Aartsen:2013jdh,Aartsen:2013bka,Aartsen:2014gkd} recently reported their observation of high-energy neutrinos in the range 30 TeV $-$ 2 PeV. The observation of neutrinos are isotropic in arrival directions. No particular pattern has been identified in arrival times. This implies that the source is not local but broadly distributed. This superheavy neutrino (LSP) might be the dark matter, distributed in the Galactic halo.
In this work we find that the thermally produced LSP can serve as a viable PeV dark matter candidate satisfying the dark matter relic density in the right ball park which also explains the IceCube excess at PeV. However, an annihilating PeV scale dark matter candidate posses two serious problems.

({\bf 1}). To maintain the correct relic density using the well-known thermal freeze-out mechanism requires a very large annihilation cross-section, which violates the unitarity bound. The s-wave annihilation cross-section of dark matter with a mass $M_{DM}$ is limited by unitarity as~\cite{Griest:1989wd},
\beq
<\sigma v>\,\, \lesssim \frac{4 \pi}{M_{DM}^2 v}.
\label{relicUni}
\eeq
The unitarity bound restrict the dark matter mass below 300 TeV~\cite{Griest:1989wd}. Also, in order to satisfy the unitarity bound, the  PeV scale dark matter model produces an overabundance $\mathcal{O}(10^{8}$) (depends on the model) of the dark matter.

({\bf 2}). The decay time of a particle is in general inversely proportional to its mass. So a PeV scale particle is generically too short-lived to be a dark matter candidate. One has to use fine-tuning~\cite{Rott:2014kfa,Fiorentin:2016avj} to stabilize the dark matter as lifetime of the DM particles has to be at least larger than the age of the Universe~\cite{Audren:2014bca,Aartsen:2014gkd}.
%

\begin{figure}[h!]
 \begin{center}
\resizebox{12cm}{!}{
 \hspace{-0cm}{\includegraphics[width=5.5in,height=0.9in, angle=0]{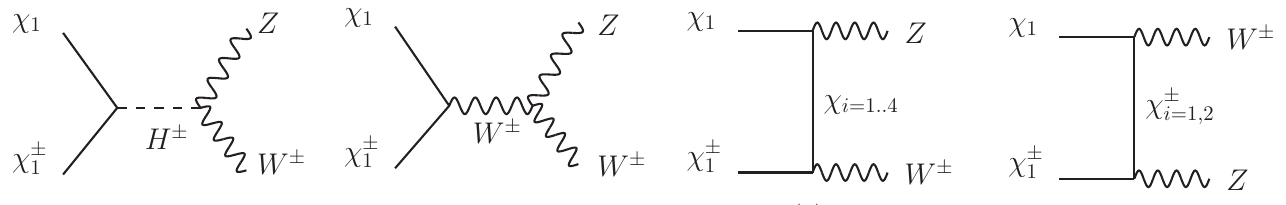}}}
 \caption{\label{fig:coanicont} \textit{Co-annihilation diagram of lightest neutralino $\chi_1$ \rm{(DM)} with the charginos $\chi^\pm_1$.}}
 \end{center}
 \end{figure}

Generally, thermal dark matter freeze-out depends on the remaining dark matter in chemical and thermal equilibrium with the SM bath, which leads to depletion of dark matter through Boltzmann suppression~\cite{Garrett:2010hd, Murayama:2007ek}. We consider the possibility that the dark matter can also decay out of equilibrium to the SM particles via $R$-parity violation. In the presence of constant s-wave effective annihilation cross-section and dark matter decay, the Boltzmann equation is given by,
\begin{equation}
\frac{dn}{dt}=-3 H n - \left\langle \sigma v\right\rangle (n^2 -n_{eq}^2) - n \sum_i \frac{1}{\tau_{i,DM}}.
\label{eq:boltz1}
\end{equation}
Here, we assume  that  the  decay  rates  of super partner of SM particles other than LSP are much faster than the rate of the expansion of the Universe, so that all the particles present at the beginning of the Universe have decayed into the lightest neutralino before the freeze-out. Therefore the density of the lightest neutralino $n$ is the sum of the density of all SUSY particles. $\tau_{i,DM}$ is the decay time for the $i^{th}$ process. In non-relativistic case $T < M_{DM}$, the equilibrium number density $n_{eq}$ is given by the classical Maxwell-Boltzmann distribution,
\begin{eqnarray}
n_{eq}&=&  \left(\frac{M_{DM} T}{2 \pi}\right)^{3/2} Exp\left(-\frac{M_{DM}}{T}\right)\,.
\end{eqnarray}
The entropy density of the Universe at temperature $T$ is $s=g_{*} T^3 \left(\frac{2 \pi^2}{45}\right)$, where parameter $g_{*}$ is the effective degrees of freedom. We use the relation of entropy density and the expansion rate of the Universe $\frac{ds}{dt}+3 H s=0$, in the eqn.~\ref{eq:boltz1}, to obtain
\begin{equation}
\frac{dY}{dt}=  s \left\langle \sigma v\right\rangle (Y^2 -Y_{eq}^2) - Y \sum_i \frac{1}{\tau_{i,DM}},
\label{eq:boltz2}
\end{equation}
where, yield $Y$ is defined as $Y\equiv\frac{n}{s}$. We solve the above equation for the decaying LSP with large decay time.
Before freeze-out annihilation term in eqn.~\ref{eq:boltz2} dominates over the exponential decay term due to very large decay time of LSP.
Therefore, integrating eqn.~\ref{eq:boltz2} between the times $t=0$ (or $T=\infty$) the start of the Universe and $t=t_f$ (or $T=T_f$) the freeze-out time, the yield $Y(t_f)$ comes out to be inversely proportional to the thermally averaged effective annihilation cross section $\left\langle \sigma v\right\rangle$ of the dark matter~\cite{Kolb}. However after freeze-out, the annihilation of LSP is no longer large enough and decay term becomes dominant. Therefore, we neglect the annihilation term compared to the decay term, and integrate the remaining equation between the freeze-out time $t_f$ and the present age of the Universe $\tau_U$, which, for $t_U >> t_f$, give the yield today as
\beq
Y(T_0) = Y(T_f) ~Exp\left(-\sum_i \frac{t_U}{\tau_{i,DM}}\right)
\eeq
which implies that the number density of LSP reduces with time.
Therefore, the relic abundance of LSP in the present Universe can be written as,
\bea
\Omega & = & \frac{s_0}{\rho_c} M_{DM} Y(T_0)=\frac{s_0}{\rho_c} M_{DM} Y(T_f) ~Exp\left(-\sum_i \frac{t_U}{\tau_{i,DM}}\right)
\label{redensity}
\eea
where $s_0\sim{\rm 2890 ~cm^{-3}}$ is the current entropy density and $\rho_c\sim {\rm 1.05\times10^{-5}} h^2 {\rm ~GeV cm^{-3}}$ is critical density of the present Universe, $h=0.72$ is a dimensionless parameter defined through the Hubble parameter $H_{0}=100\,h \, {\rm km \, sec^{-1} Mpc^{-1}}$. 
This implies that the relic density is related to the annihilation cross-section and the decay life time as 
\beq
\Omega h^2 ~\simeq~ \frac{9.62 \times 10^{-28} \,[\text{cm}^3\text{sec}^{-1}]}{\left\langle \sigma v\right\rangle}~Exp\left(-\sum_i \frac{t_U}{\tau_{i,DM}}\right)
\label{relic1}
\eeq
Using particles spectrum (see table~\ref{tab2}) in {\tt micrOMEGAs}, we get
\beq
\left\langle \sigma v\right\rangle = 6.78 \times 10^{-26} ~\text{cm}^3\text{sec}^{-1}\,.
\label{nereli}
\eeq
%
The main contribution to the DM effective annihilation cross-section comes from the co-annihilation channel (see Fig.~\ref{fig:coanicont}) $\chi_{1}\chi_{1}^\pm\rightarrow Z W^\pm$ which is consistent with unitarity bound~\cite{Griest:1989wd} and gives the bino-type DM density $\Omega h^2 \simeq 0.0142 \,\rm{Exp}\left(-\sum_i \frac{t_U}{\tau_{i,DM}}\right)$. The joint results of CMB observations from WMAP and Planck Collaboration~\cite{Ade:2013zuv} give $\Omega h^2 = 0.1198 \pm 0.0026$. It has been shown that if a subdominant fraction of DM is decaying~\cite{Anchordoqui:2015lqa,Berezhiani:2015yta}, then it can resolve the conflict between $\sigma_8$ and $H_0$ which exists in $\Lambda$CDM model. Berezhiani et. al. have shown that if $\sim 5\%$ of the dark matter has a decay lifetime of $1.6\times10^{16}$ seconds resolves the $\sigma_8-H_0$ conflict~\cite{Berezhiani:2015yta}. For the PeV DM Anchordoqui et. al. have shown that $\sim 5\%$ of DM of mass scale $76$ PeV and lifetime $6\times10^{15}$ seconds is required for explaining the IceCube events~\cite{Anchordoqui:2015lqa}. In this paper, we assume the lifetime of DM decay to leptons via the $R$-parity violation $LLe^c$ operator is $8.9\times10^{28}$ seconds which gives $11\%$ (i.e. $\Omega h^2 \simeq 0.0142$) of the DM relic density and is consistent with the structure formation. The neutrino flux is obtained by this decay channel to give the PeV neutrino flux required to explain the IceCube's HESE data~\cite{Schneider:2019ayi, Aartsen:2013jdh,Aartsen:2013bka,Aartsen:2014gkd}.
We will explain this result in the subsection~\ref{IceCubeFit} in more detail.

\begin{figure}[h!]
 \begin{center}
 {\includegraphics[width=2.8in,height=1.8in, angle=0]{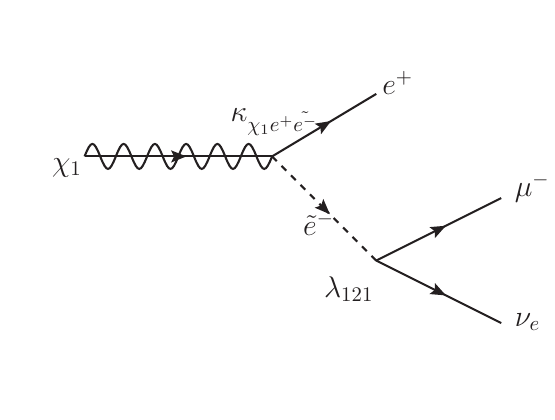}}
 \hskip 0.4cm
 {\includegraphics[width=2.8in,height=1.8in, angle=0]{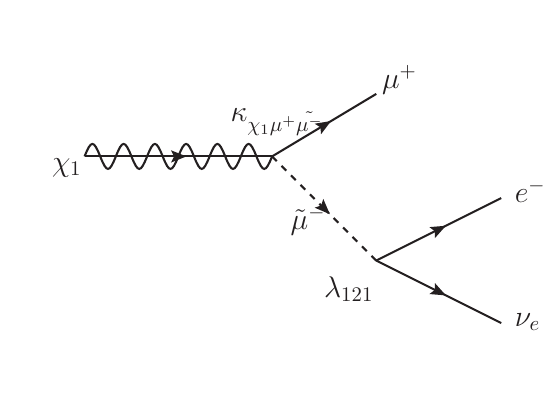}}
 \caption{\label{fig:relicflux} \textit{Decay diagrams of neutralino via lepton number violating coupling which produces the Ice-cube flux.}}
 \end{center}
 \end{figure}

The presence of $R$-parity violating couplings~\ref{SP2} can explain the high energy neutrino events at IceCube. We take only one non-zero dimensionless trilinear $R$-parity violating couplings namely $\lambda_{121}$ coefficient of $L_e L_\mu e^c$. $\lambda_{121}$ helps in producing the neutrino flux through the decay of the Neutralino (see Fig.~\ref{fig:relicflux}).
We find the decay width for the dark matter decaying into $e^+ \mu^- \nu_e$ and $e^- \mu^+ \nu_e$~\cite{Baltz:1997gd} as
\begin{equation}
\Gamma (\chi_1 \rightarrow e^+ \mu^- \nu_e) = \frac{ \kappa_{{\chi_1} e^+ \tilde{e}^-}^2 {\lambda_{121}^2} M_{\chi_1}^5 }{(8 \pi )^3 M_{\tilde{e}^-}^4}~~~~~~{\rm and}~~~~~~\Gamma (\chi_1 \rightarrow e^- \mu^+ \nu_e) = \frac{ \kappa_{{\chi_1} \mu^+ \tilde{\mu}^-}^2 {\lambda_{121}^2} M_{\chi_1}^5 }{(8 \pi )^3 M_{\tilde{\mu}^-}^4}
\end{equation}
where coupling $\kappa_{{\chi_1} e^+ \tilde{e}^-}=\kappa_{{\chi_1} \mu^+ \tilde{\mu}^-}\sim \sqrt{2} g_2$~\cite{Drees:2004jm,Martin:1997ns}, $M_{\chi_{1}}\simeq 6.3$ PeV and $M_{\tilde{e}^-,\tilde{\mu}^-}\simeq 14.36$ PeV (see the Table~\ref{tab2}). For the lepton number violating coupling $\lambda_{121}\approx 2.82 \times 10^{-28}$, we get the decay time for this channel $\tau_{\nu,DM} \approx 8.9 \times 10^{28} {\rm sec}$. It is to be noted that we need very tiny $\lambda$'s to explain the IceCube HESE data. It indeed has the fine-tuning issue and the dynamical explanation lies somewhere else. To explain IceCube HESE data in terms of decaying dark matter, fine-tuning actually seems to be the most ``natural" option~\cite{Carena:1998gd,Anchordoqui:2006wc,Dev:2016uxj,Becirevic:2018uab}.

\subsection{Fitting the IceCube data}
\label{IceCubeFit}
In this subsection, we fit the flux of neutrinos observed at IceCube from the decay of the netralino LSP.
The total contribution to the neutrino flux from the atmospheric background and the atrophysical sources along with the galactic (G) DM halo and extragalactic (EG) DM is given by,
\beq
\frac{d\Phi_{tot}}{dE_\nu} = \frac{d\Phi_{Atm}}{dE_\nu} + \frac{d\Phi_{astro}}{dE_\nu} + \frac{d\Phi_{G}}{dE_\nu} + \frac{d\Phi_{EG}}{dE_\nu}\label{NuFluxx}
\eeq
Following the analysis as in Ref.~\cite{Roland:2015yoa} for our mSUGRA model parameters, we compute the number of neutrino events as a function of deposited energy
\beq
N_{bin} = T_{E} \int_{E_{min}^{bin} (E_{dep})}^{E_{max}^{bin}(E_{dep})} A(E_\nu) \frac{d\Phi_{tot}}{dE_\nu} \, dE_{dep}(E_\nu) ,
\eeq
where $T_{E}=2635$ is the total exposure time, $E_{dep}$ is the deposited energy in the laboratory frame and $A(E_\nu)$ is the neutrino effective area for particular lepton flavor~\cite{Palomares-Ruiz:2015mka}. We sum over all the neutrino flavors. Due to low statistical data points at IceCube, it is acceptable to assume that the two energies coincide, i.e., $E_{dep} \simeq E_\nu$~\cite{Aisati:2015vma}.

 \begin{figure}[h!]
 \begin{center}
 \includegraphics[width=3.8in,height=2.8in, angle=0]{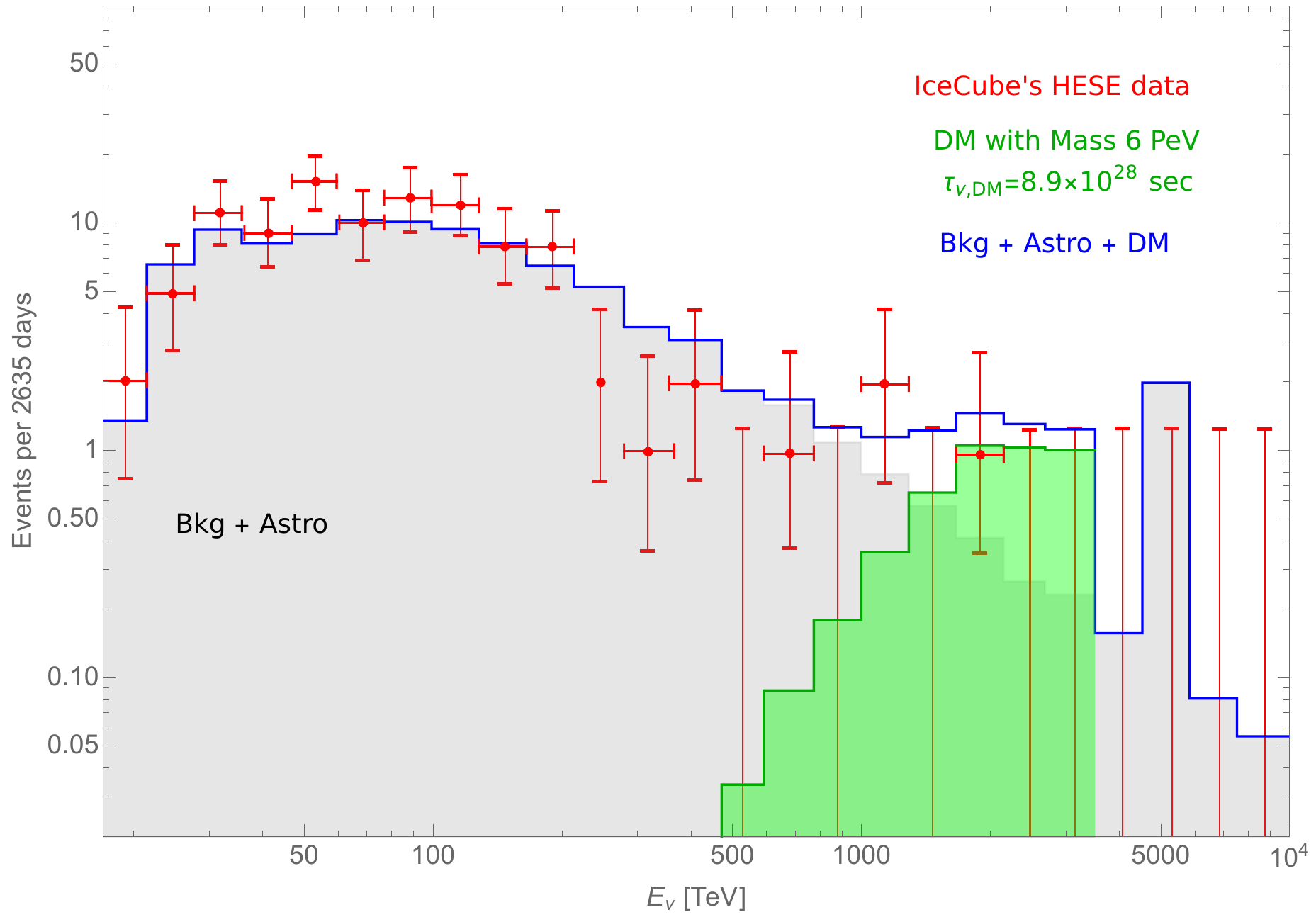}
 \caption{\textit{Normalized neutrino events distribution is shown. The contributions to PeV neutrino flux from the galactic and the extragalactic DM decay is shown by the green region. The total flux from all the sources is shown by the the blue line.}}\label{NuFlux}
 \end{center}
 \end{figure} 

We have extracted the atmospheric background data from Ref.~\cite{Aartsen:2014muf} and added it to the data from astrophysical sources which come mainly due to the decay of highly energetic pions. The source of these pions could be known sources like active galactic nuclei or the Supernova remnants. We use the power-law flux $ E_\nu^2 \frac{d\Phi_{astro}}{dE_\nu} = \Phi_0 (E_\nu)^{-\gamma}$ with $\Phi_{0}=2.2\times10^{-8}\,\, {\rm cm^{-2} sec^{-1} sr^{-1}}$ and $\gamma=0.8$~\cite{Aisati:2015vma} to fit the data from the astrophysical sources. Both the galactic and the extragalactic neutrino flux contributions are important to explain the excess neutrino flux around 1-2 PeV ~\cite{Roland:2015yoa,Borah:2017xgm,Mileo:2016zeo,Aisati:2015vma,Palomares-Ruiz:2015mka,Shakya:2015tza,Dev:2016qbd}.
We take NFW (Navarro, Frenk, and White) DM density profile to account for the galactic dark matter contribution, where the neutrino energy spectrum $\frac{dN}{dE_\nu}$ evaluated at our model parameter values has been used~\cite{Esmaili:2012us}. We present the neutrino events distribution in Fig.~\ref{NuFlux}. The contribution to the neutrino flux from the DM decay can explain the PeV excess at IceCube.

\vspace{-.2cm}
\section{Conclusions} \label{sec:concl}
In this paper we have constructed a Starobinsky like Higgs-sneutrino model of plateau inflation from supergravity $D$-term in MSSM fields, and $6$ PeV bino-type dark matter gives the observed flux of PeV neutrino events at IceCube HESE. A subdominant fraction ($\sim 11\%$) of the DM relic density and its decay to neutrinos is obtained by choosing the couplings of the $R$-parity violating operators. The SUSY breaking is obtained by Polonyi field which sets the scale of soft SUSY parameters $m_0$, $m_{\frac{1}{2}}$, $A_0$ and $m_{\frac{3}{2}}$ as a function of the parameters of the model. By running the RGEs, we show the low energy (PeV scale) spectrum of SUSY particles and show that the model can accommodate the $\sim$125 GeV Higgs. The SUSY spectrum so obtained modifies the RGEs above the PeV scale and gives the coupling constant unification at the GUT scale, however such a gauge coupling unification can be achieved for a large range of SUSY-breaking scales. We find that there is a degeneracy in the values of inflation model parameters $(\gamma, \beta, \zeta))$ in predicting the correct CMB amplitude. However $\tan(\beta)=1.8$, fixed from SUSY breaking, removes this degeneracy and provide a unique solution for inflation, and brings the explanation for dark matter, PeV neutrinos and inflation within the same model setup.

\section*{Data Availability}
The data used to support the findings of this study are available from the corresponding author upon request.

\section*{Conflicts of Interest}
The author declares that there are no conflicts of interest regarding the publication of this paper.

\section*{Acknowledgements}
G.K.C. would like to thank University Grants Commission, Govt. of India to provide financial support via Dr. D. S. Kothari Postdoctoral Fellowshipp (Grant No.: BSR/PH/2017-18/0026). 
N. K. acknowledges private communications with Florian Staub.

\providecommand{\href}[2]{#2}\begingroup\raggedright\endgroup

\end{document}